\documentclass[]{JHEP3}
\usepackage{graphicx}
\JHEPspecialurl{http://jhep.sissa.it/JOURNAL/JHEP3.tar.gz}

\usepackage{amsmath}
\usepackage{epsfig,multicol,bbm}

\def\slash#1{\setbox0=\hbox{$#1$}#1\hskip-\wd0\hbox to\wd0{\hss\sl/\/\hss}}

\newcommand{\newc}{\newcommand}
\newc{\gsim}{\lower.7ex\hbox{$\;\stackrel{\textstyle>}{\sim}\;$}}
\newc{\lsim}{\lower.7ex\hbox{$\;\stackrel{\textstyle<}{\sim}\;$}}
\newc{\gev}{\,{\rm GeV}}
\newc{\mev}{\,{\rm MeV}}
\newc{\ev}{\,{\rm eV}}
\newc{\kev}{\,{\rm keV}}
\newc{\tev}{\,{\rm TeV}}
\def\ln{\mathop{\rm ln}}

\newc{\mz}{M_Z}
\newc{\mpl}{M_*}
\newc{\mw}{m_{\rm weak}}
\newc{\nr}[1]{N^c_R{}_{#1}}
\usepackage{amsmath}
\def\beq{\begin{equation}}
\def\eeq{\end{equation}}
\def\bea{\begin{eqnarray}}
\def\eea{\end{eqnarray}}
\def\bitem{\begin{itemize}}
\def\eitem{\end{itemize}}
\newc{\ie}{{\it i.e.}}          \newc{\etal}{{\it et al.}}
\newc{\eg}{{\it e.g.}}          \newc{\etc}{{\it etc.}}
\newc{\cf}{{\it c.f.}}

\newcommand{\lang}{\mathcal{L}}

\newcommand{\half}{\frac{1}{2}}

\def\inv{^{\raise.15ex\hbox{${\scriptscriptstyle -}$}\kern-.05em 1}}
\def\lbar{{\lower.35ex\hbox{$\mathchar'26$}\mkern-10mu\lambda}} 

\def\to{\rightarrow}

\let\<=\langle
\let\>=\rangle

\let\+=\uparrow

\newcommand{\pd}[2]{\frac{\partial #1}{\partial #2}}

\newcommand\fverb{\setbox\fverbbox=\hbox\bgroup\verb}
\newcommand\fverbdo{\egroup\medskip\noindent%
			\fbox{\unhbox\fverbbox}\ }
\newcommand\fverbit{\egroup\item[\fbox{\unhbox\fverbbox}]}
\newbox\fverbbox

\title{Axion-Assisted Electroweak Baryogenesis}

\author{Nathaniel Craig\\
	Institute for Theoretical Physics, Stanford University, Stanford, CA 94306, USA\\
	E-mail: \email{ncraig@stanford.edu}}

\author{John March-Russell\\
	Rudolf Peierls Centre for Theoretical Physics, University of Oxford, 1 Keble Road, Oxford, OX1 3NP, UK\\
	E-mail: \email{jmr@thphys.ox.ac.uk}}

\preprint{SU-ITP-10-18 \\ OUTP-10-10P}

\abstract{We consider a hidden-valley gauge sector, $G$, with strong coupling scale $\Lambda\sim\tev$ and CP-violating topological parameter, $\theta$, as well as a new axion degree of freedom which adjusts $\theta$ to near zero in the current universe.  If the $G$-sector couples to the Standard Model via weak-scale states charged under both, then in the early universe it is possible for the CP-violation due to $\theta$ (which has not yet been adjusted to zero by the hidden axion) to feed in to the SM and drive efficient baryogenesis during the electroweak (EW) phase transition, independent of the effectively small amount of CP violation present in the SM itself.  While current constraints on both the new axion and charged states are easily satisfied, we argue that the LHC can investigate the vast majority of parameter space where EW-baryogenesis is efficiently assisted, while the hidden axion should comprise a significant fraction of the dark matter density.   In the supersymmetric version, the ``messenger'' matter communicating between the SM- and $G$-sectors naturally solves the little hierarchy problem of the MSSM.   The connection of the hidden scale and masses of the ``quirk''-like messengers to the EW scale via the assisted electroweak baryogenesis mechanism provides a reason for such new hidden valley physics to lie at the weak scale.}


\begin{document} 

\maketitle

\vfill\eject
\section{Introduction}

The string landscape of vacua allows a rich variety of sectors hidden, or partially hidden, from
the Standard Model (SM) which can lead to experimentally or observationally accessible new phenomena.   Recent discussions have focussed particularly on the superlight axion-like states ubiquitous in string compactifications---the ``axiverse''  \cite{Arvanitaki:2009fg}---or weak-scale supersymmetric states, such as photini from hidden
U(1)'s \cite{Arvanitaki:2009hb}, or goldstini from independent supersymmetry breaking sectors \cite{Cheung:2010mc}, both of which 
can strongly change LHC phenomenology, astrophysical observations, and early universe cosmology.  

In this paper we turn to another way in which a mildly-sequestered hidden sector, now possessing a heavy axion, may impact the physics and phenomenology of the SM.   The essential idea is simple to state:  Consider a hidden sector gauge theory analogous to QCD with its own additional CP-violating topological $\theta$-parameter, as well as a (heavier than QCD-) axion degree of freedom which automatically adjusts this theta angle to near zero in the current universe.  We argue that if this extra sector couples to the SM, say via TeV-scale bi-fundamental states, then, in the early universe, it is possible for the CP-violation due to $\theta$ (which has not yet been adjusted to zero by the hidden axion) to feed in to the SM and drive {\it efficient} baryogenesis during the electroweak phase transition.\footnote{Assuming, as usual, that the EWPT is sufficiently out-of-equilibrium.  Here our focus is on the alterations to standard EW-baryogenesis due to the hidden sector $\theta$-term and axion, and we take the view that the new sector and couplings to the SM  are ``retrofitted" to one of the many TeV-completions of the SM with sufficiently out-of-equlibrium dynamics (not necessarily only a strong 1st order EW phase transition).  We will return
to the detailed investigation of EWPT out-of-equlibrium dynamics with axion-assistance in a later work.}   In contrast
to the CP violation present in the SM itself, which even for large CKM phase is very small due to suppression by powers
of the squared differences of quark Yukawa couplings, the physical CP violation induced at the time of the EWPT can be large
in our mechanism.    Moreover, the stringent present
experimental constraints on CP violation in extensions of the SM are simply and easily met by this construction as the
hidden-sector $\theta$-angle is naturally relaxed to near zero at current temperatures.   

One may ask why a similar mechanism is not operable in an even simpler extension of the SM, the SM with QCD $\theta$-term
now relaxed to very near zero by the conventional Peccei-Quinn-Weinberg-Wilczek QCD axion \cite{Peccei:1977hh,Weinberg:1977ma,
Wilczek:1977pj}: Similar to our model the
QCD axion has not yet relaxed to its $\theta_{QCD}$-cancelling minimum at the time of the EWPT, so it seems that large
CP violation is available in this situation as well.  This is not the case, however, as the $\theta$-term is topological and
only leads to physical effects via topologically non-trivial QCD field configurations, which at temperatures $T\gg \Lambda_{QCD}$
have large action and are suppressed \cite{Kuzmin:1992up}.  This leads to an effective size of CP-violation suppressed by $\sim (\Lambda_{QCD}/T_{EW})^{13} \ll 1$, even for a bare $\theta_{QCD} \sim 1$.    

Our scenario is distinctive in that the strong dynamics in the hidden sector must lie close to or somewhat above the EW scale, and the resulting CP-violation is communicated to the SM by new massive matter that must be charged under both the SM and the hidden gauge group, and -- for successful baryogenesis -- cannot be parametrically much heavier than the EW scale. Such confining hidden sectors with weak-scale matter and couplings to the SM have attracted a considerable amount of study on account of their novel phenomenology (see, e.g., \cite{Strassler:2006im,Han:2007ae, Burdman:2006tz, Kang:2008ea, Kilic:2009mi, Juknevich:2009ji, Juknevich:2009gg}); our viewpoint is that the connection to electroweak baryogenesis provides a  {\it reason} for such new physics to lie at the weak scale.
These features lead in turn to experimental signatures of our scenario at the LHC, and associated constraints from precision
electroweak tests and direct searches at LEP/Tevatron.   The constraints are easily satisfied, and leave a large range of parameter
space which is accessible to investigation at the LHC.   An amusing feature of a supersymmetric version of our mechanism is that the 
``messenger'' matter communicating between the SM and hidden sectors naturally solves the little hierarchy problem of the MSSM
for the range of parameters for which EW-baryogenesis is effectively assisted.   Even though the hidden sector axion is heavier than the QCD axion by a factor $\sim (v_{EW}/\Lambda_{QCD})^2 \sim 10^6$, the prospects for detecting the new axion
itself are similar to that of the standard QCD axion for equivalent axion decay constant $f_G$.   Intriguingly, as we show in detail below,
efficient axion-assisted electroweak baryogenesis (AAEB) favours, $f_G \sim 10^{16}\gev$, the natural string value of the axion decay constant\cite{Svrcek:2006yi,Arvanitaki:2009fg}, and disfavours the traditional axion window, $10^9  \gev \lsim f_G \lsim 10^{12} \gev$.  Thus, as $f_G$ is in the anthropic range requiring a mildly fine-tuned initial misalignment angle, a further natural consequence of our mechanism is that the hidden axion comprises a significant fraction of the dark matter density.

The setup of our paper is as follows: In Section \ref{sec:ba} we review the relation between the observed baryon asymmetry and a class of CP-odd effective operators. In Section \ref{sec:model} we present a simple model, based on a confining hidden sector with an axion-like coupling and matter fields charged under the SM, and show that confinement near the electroweak scale may naturally give rise to the required degree of CP violation during electroweak baryogenesis.  Such additional degrees of freedom are naturally subject to a variety of experimental and cosmological constraints, which we consider in Section \ref{sec:constraints}. The presence at the weak scale of new strongly-interacting fields charged under the SM raises the prospect of various LHC signatures and other phenomenological consequences, to which we turn in Section \ref{sec:pheno}. In Section \ref{sec:susy} we briefly consider a supersymmetrized version of the model, for which the hidden sector matter content naturally solves the little hierarchy problem. We conclude in Section \ref{sec:conc}, and reserve for the Appendix a detailed discussion of the axion mass and evolution.

\section{Generating the Baryon Asymmetry \label{sec:ba}}

A particularly simple way of generating a sufficient amount of CP violation is to incorporate new physics giving rise, at
low energies, to an effective (and CP-odd) irrelevant operator of the form \cite{Shaposhnikov:1987pf}\footnote{In this
Section we follow closely the discussion of the QCD case presented in Ref.\cite{Kuzmin:1992up}.}
\beq \label{eqn:cplang}
\lang_{CP} = \frac{g^2}{32 \pi^2} W_{\mu \nu} \tilde W^{\mu \nu} \Phi(T,H)
\eeq
where $W$ is the $SU(2)$ gauge field strength and $\Phi$ is some temperature-dependent functional of the Higgs field(s).  The functional $\Phi$ acts like a type of $T$-dependent, and thus, in the early universe, time-dependent, effective ``axion" for $W \tilde W$, driving CP violation into the SM sector during the EW transition and thus assisting EW baryogenesis.   

Using the anomaly equation, $\frac{g^2}{32 \pi^2} W_{\mu \nu} \tilde W^{\mu \nu} = \partial_\mu j^\mu_{CS}$, an interaction of the above form may be expressed, by integrating by parts and focusing on the time component, as a chemical potential for the Chern-Simons number density for $SU(2)$, which in turn biases the change in baryon number $\lang_{CP} = j^0_{CS} \partial_0 \Phi =
n_{CS} d\Phi/dt$.
Hence the effective Lagrangian arising from irrelevant operators of the form Eq.(\ref{eqn:cplang}) looks like a chemical potential for Chern-Simons number density, where the effective chemical potential is given by $\mu_{CS} = \dot{\Phi}$.  It is then relatively straightforward to evaluate the contribution to CP violation in baryogenesis for various forms of the functional $\Phi$. The size of CP violation ultimately depends on both the size of the coefficient in $\Phi$ and the degree of time variation during baryogenesis. In general, if the only time-variation in $\Phi$ comes from the time-variation of $T$, then $\mu_{CS}$ is suppressed by a factor $T/M_P$ encoding the rate of Hubble expansion. It's only during a first-order phase transition that we expect more rapid time variation of the quantity $\Phi$, and hence an unsuppressed chemical potential for $n_{CS}$.

Although there are a variety of nonequilibrium phenomena that may lead to baryogenesis, we will focus here on baryogenesis during the electroweak phase transition \cite{Kuzmin:1985mm}, under the assumption that the phase transition is first-order and sufficiently strong.\footnote{For excellent reviews, see \cite{Cohen:1993nk, Trodden:1998ym, Riotto:1999yt}.}  In principle, the phase transition proceeds through bubble nucleation, and the generation of baryon number asymmetry occurs only in the bubble wall when $\dot{\Phi}$ is significant. In practice, however, it is difficult to make any precise quantitative statements about a nucleation phase transition without extensive numerical simulation. One must understand not only the details of the phase transition and bubble wall propagation, but also particle transport at the phase boundary in the presence of CP violation.\footnote{For a recent discussion see, e.g., \cite{Cirigliano:2009yt} and references therein.}  Qualitative estimates may be made if we consider a spinodal decomposition phase transition, in which the scalar field rolls {\it uniformly} to the true vacuum; such a transition gives a spatially uniform, but time-varying, phase transition. The assumption is that the results should be similar to those of a nucleation phase transition, as Lorentz invariance in principle relates processes with time-varying fields to those with space-varying fields. The resulting expression for baryon asymmetry is expected to reflect the correct parametric dependence on microscopic parameters, up to $\mathcal{O}(1)$ coefficients. We emphasize, however, that this simplifying assumption is made strictly for the purpose of understanding the correct parametric dependence of baryon asymmetry upon the microphysical parameters of our model; it is reasonable insofar as the magnitude of CP violation is independent from the details of the phase transition.

Assuming we have some free energy difference $\Delta F$ between neighbouring minima, and a rate $\Gamma_a$ for fluctuations between neighbouring minima (in the absence of bias, as we assume the bias is a small perturbation), the master equation for baryon number is \cite{Dine:1989kt}
\beq
\frac{d n_B}{dt} = - 3 \frac{\Gamma_a}{T} \Delta F
\eeq
We can write this in a somewhat more useful form via $\Delta F = n_f \pd{F}{B}$ (where $n_f =3$) to obtain
\beq
\frac{d n_B}{dt} = - 3 \frac{\Gamma_a \mu_B}{T} = - \frac{\Gamma_a \mu_{CS}}{T}
\eeq
The number density of baryons created during the phase transition is then given by 
\beq
n_B = \frac{n_f}{T} \int_0^\infty dt \, \Gamma_a(t) \mu_{CS}(t) .
\eeq
The crucial part, of course, rests in accurately evaluating the chemical potentials biasing baryon number production, as well as the time evolution of rates.

The transition rate for fluctuations between neighbouring minima depends entirely whether electroweak symmetry is broken. In the unbroken phase, sphaleron transitions are unsuppressed, while in the broken phase there arises the usual exponential suppression.  That the transition rates in the symmetric and broken phases are, respectively,\cite{Cohen:1993nk}
\bea
\Gamma_a &\simeq& 30 (\alpha_w)^5  T^4 \hspace{2.8cm} \text{(symmetric)} \\
\Gamma_a &\simeq& (\alpha_w T)^{-3} m_W^7 e^{-E_{sph} / T} \hspace{1cm}\text{(broken)}
\eea
Clearly the contribution from the symmetric phase dominates. In the spinodal decomposition transition, we  may assume that these values interpolate smoothly. Treating $\Gamma_a$ as a step function and $T$ as essentially constant during the phase transition, we can estimate the integral to find a total baryon asymmetry
\beq 
\Delta \equiv \frac{n_B}{s} \simeq \frac{675}{\pi g_*} n_f \alpha_w^5 \frac{T_i^3}{T_f^3} \delta \Phi
\label{eqn:baryonasymmetry}
\eeq
where $T_i, T_f$ are the temperatures before and after the phase transition; $s = (2 \pi^2 g_* / 45) T^3$, and $\delta \Phi \equiv \Phi(T_i,H = g_W T_i) - \Phi(T_i, H = 0)$. This estimate should be compared to the observed value, $n_B/s \simeq 10^{-10}$. Although this expression has been obtained for a spinodal decomposition transition with a number of simplifying assumptions, it is parametrically similar to the analogous expression for bubble nucleation; see, e.g., \cite{Joyce:1994zn, Joyce:1994zt}. In the case of bubble nucleation, the expression $(T_i/T_f)^3$ is typically replaced by $\sim (m_t / T_c)^2 (m_h/T_c),$ but the parametric dependence on $\delta \Phi$ is unaltered.  It is then a fairly straightforward procedure to compute the size of CP violation $\delta \Phi$ given a particular functional $\Phi$, to which we will now turn.

\section{A Toy Model \label{sec:model}}

Consider now an additional nonabelian gauge sector $G$ with fundamental and antifundamental fermions carrying vector-like charges under SM $SU(2)_L \times U(1)_Y$. We will assume that these fields are uncharged under $SU(3)_C$; this assumption may be relaxed, which leads to more stringent constraints on vector masses and significantly altered collider phenomenology, among other things. The matter content is such that the group $G$ is asymptotically free and confines at some scale $\Lambda_G$; such additional confining sectors have been the subject of considerable study in recent years \cite{Strassler:2006im,Han:2007ae,Kang:2008ea, Kilic:2009mi,Juknevich:2009ji}. We will henceforth refer to the gauge sector $G$ as $G$-color, and the fermions charged under $G$ as $G$-quarks or $G$-fermions. While there are a variety of candidate models that exhibit the same qualitative features, among the simplest such candidates is $G = SU(N)_G$ with the matter content shown in Table \ref{tab:matter}.

\TABLE[r]{
\begin{tabular}{|c|c|c|c|}
\hline
& $SU(N)_G$ & $SU(2)_L$ & $U(1)_Y$ \\ \hline
$Q$ & $\Box$ & 2 & $Y_Q$ \\
$\overline{Q}$ & $\overline{\Box}$ & 2 & $-Y_Q$ \\
$U$ & $\Box$ & 1 & $1/2+Y_Q$ \\ 
$\overline U$ & $\overline{\Box}$ & 1 & $-1/2-Y_Q$ \\  
\hline
\end{tabular}
\label{tab:matter}\caption{Model matter content.  Here $Y_Q$ is an arbitrary hypercharge assignment.}
}

The added $G$-fermion matter content is free of gauge and gravitational anomalies. We may introduce vector masses for the $G$-fermions, as well as appropriate couplings to the SM Higgs
field $H$:
\beq
\lang_G \supset -  \mu_Q Q \overline Q - \mu_U U \overline U - \lambda H^\dag Q \overline U -  \lambda' H \overline Q U + \text{ h.c.}
\eeq
The $G$-fermions acquire mass from both their vector masses and from electroweak symmetry breaking (EWSB) via their coupling to the Higgs. The mass eigenstates consist of two  Dirac $G$-fermion-antifermion pairs of electric charge $\pm (Y_Q + 1/2)$ and one Dirac $G$-fermion-antifermion pair of charge $\pm(Y_Q - \half)$. The mass of the latter $G$-fermion pair is simply $\mu_Q$, while the mass matrix for the charge $\pm(Y_Q+1/2)$ $G$-fermions is 
\beq
\mathcal{M} = \left( \begin{array}{cc}
\mu_Q & \frac{1}{\sqrt{2}} \lambda v(T) \\
\frac{1}{\sqrt{2}} \lambda' v(T) & \mu_U
\end{array} \right)
\eeq
where have emphasized the temperature-dependence of the EWSB Higgs vev which will be important to our mechanism.
In the simplest case with $\mu_Q = \mu_U = \mu$, the mass eigenvalues for the three Dirac fermions are
then \beq
m_Q(T)= \mu, \; \mu \pm \sqrt{\frac{\lambda \lambda'}{2}} v(T). 
\label{eqn:mQT}
\eeq
In Section \ref{sec:susy}, in the context of a supersymmetric version of our toy model, we will argue that the vector-like mass terms, $\mu_Q, \mu_U$,  arise in the same fashion as the SUSY-Higgs $\mu$-term, thus justifying both their link to the EW scale, and our notation.  At present we will just take them to be free parameters of the toy model.

Of course, this matter content spoils na\"{i}ve unification of SM gauge couplings. Although conventional gauge coupling unification is by no means a necessary ingredient, it is possible to restore gauge coupling unification by adding additional $G$-quarks charged under $SU(3)_C$, with suitable SM charge assignments filling out complete $\bf{5} + \overline{\bf{5}}$ multiplets under $SU(5)_{GUT}$. We will assume that such additional $G$-quarks, if present, have vector masses $\gtrsim 1$ TeV, and so do not contribute meaningfully to the low energy phenomenology.

The $G$-color gauge sector is asymptotically free, and confines in the IR at a scale
\beq
\Lambda_G = \mu_0 e^{-8 \pi^2 /b_G g_{G,0}^2}
\eeq
given an initial value of the $G$ gauge coupling $g_{G,0}$ at a scale $\mu_0$. In this case $b_G = \frac{11}{3} T(G) - \frac{4}{3} N_f T(R) = \frac{11}{3} N - 2 $. We will be interested in a range of possible values of $\Lambda_G$ relative to $m_Q$. In any case, since the $G$-fermions are vectorlike under the SM, a potential condensate of $G$-fermions leaves electroweak symmetry unbroken.

It is completely consistent with the symmetries of the theory to include also a CP-odd $\theta$-term for the $G$-color gauge group,
\beq
\lang_G \supset - \frac{\alpha_G \theta_G}{8 \pi} G_{\mu \nu} \tilde G^{\mu \nu}.
\eeq
Such $\theta$ terms are generically present in nonabelian hidden sectors, some consequences of which were considered in \cite{Cassel:2009pu}. Of course, the bare angle $\theta_G$ may be shifted by rotations of the $G$-quark mass matrix, $\theta_G \to \theta_G + \arg \det \mathcal{M} \equiv \overline \theta_G$; for clarity we will leave this redefinition implicit. While the analogous CP angle in the SM is constrained to satisfy $|\overline \theta_{QCD}| < 10^{-9}$, there are no direct constraints on the value of $\theta_G$.\footnote{Notice that this may change if the $G$-quarks are charged under QCD, in which case there arises an effective contribution to $\theta_{QCD}$ of order $\sim \frac{\alpha_G}{8 \pi} \frac{\langle G \tilde G \rangle}{m_Q^4} \theta_G$.}

The CP angle for the $G$-color sector becomes particularly interesting in the event that there is also a pseudo-Goldstone axion $a_G$ coupling to $G$-color via the usual operator
\beq
\lang_G \supset \frac{\alpha_G}{8 \pi} \frac{a_G}{f_G} G \tilde G
\eeq
where $f_G$ is the scale of PQ symmetry breaking for the $G$-axion. Although such  axions are not necessary ingredients of a nonabelian hidden sector (unlike the SM axion, required to explain the smallness of $\overline \theta_{QCD}$), they arise quite naturally in many string compactifications \cite{Svrcek:2006yi}.\footnote{This mechanism for enhancing CP violation during baryogenesis  may also work in a hidden sector without an axion, provided a nonzero vacuum expectation value for the $G$-gluon condensate $\langle G \tilde G \rangle$, but this lacks the tidy relaxation of $\theta_G$ at late times. Moreover, additional axions are a natural consequence of string compactification, and therefore not unexpected in this context.}  The axion acquires a zero-temperature mass from confinement of $G$ of order $m_a^2 \sim \frac{m_Q \Lambda_G^3}{f_G^2},$ which may also accumulate finite-temperature corrections. There also may be contributions to the $G$-axion mass coming from string instantons, but these are subdominant to the mass coming from $G$-color confinement if the string-compactification is such that the QCD axion solves the usual strong-CP-problem, as discussed in \cite{Arvanitaki:2009fg}.

In general, the $G$-axion obtains a nonzero vacuum expectation value; as with the QCD axion, its expectation value may naturally be of order $\langle a_G \rangle \sim f_G$ when its PQ symmetry is broken, leading to a nonzero angle $\theta_G.$ When the amplitude of the $G$-axion is nonvanishing, there is a condensate of $G \tilde G$. We can parameterize this as usual via 
\beq
\frac{\alpha_G}{8 \pi} \langle G \tilde G \rangle = m_a^2(T) f_G^2 \sin \theta_G
\eeq
where $m_a(T)$ is appropriately the temperature-dependent axion mass.

Below the scale $m_Q$, integrating out the massive $G$-fermions gives rise to a variety of effective operators, the most important of which (for our purposes) is the coupling to the $SU(2)_L$ $W \tilde W$ term: 
\beq
\lang_{eff} \sim \frac{\alpha_W \alpha_G}{64 \pi^2} \frac{1}{m_Q^4} W_{\mu \nu} \tilde W^{\mu \nu} G_{\mu \nu} \tilde G^{\mu \nu}
\eeq
Whenever a nonzero amplitude for the $G$-axion gives a nonvanishing expectation value to $G \tilde G$, this produces an effective $\theta$ angle for $SU(2)_L$. The result is a temperature-dependent (and thus time-dependent) effective operator
\beq \label{eqn:su2theta}
\lang_{eff} \sim \frac{g^2}{32 \pi^2} W_{\mu \nu} \tilde W^{\mu \nu} \left[ \sum_i \frac{1}{m_{Q,i}^4(T)} m_a^2(T) f_G^2 \sin \theta_G \right].
\eeq
This is precisely of the form Eq.(\ref{eqn:cplang}).  Thus in this case we 
obtain a chemical potential for CS-number of order
\beq
\mu_{CS} \simeq \sin \theta_G f_G^2 \frac{d}{dt} \left( \sum_i \frac{m_a^2(T)}{m_{Q,i}^4(T)} \right),
\eeq
and can identify the change $\delta\Phi$ appearing in the expression for the final baryon asymmetry, Eq.(\ref{eqn:baryonasymmetry}) to be
\beq
\delta\Phi(T,H) \sim \delta \left(\sum_i \frac{m_a^2(T)}{m_{Q,i}^4(T)}\right)  f_G^2 \sin \theta_G .
\label{eqn:phidef}
\eeq

If the time dependence of the $G$-color fermion and axion masses is significant around the time of the electroweak phase transition, the $G$-color sector will feed a large amount of CP violation to the SM. The resulting contribution to CP violation during electroweak baryogenesis will, in general, be unsuppressed by factors of $H/T$; both $m_Q$ and $m_a$ depend on the Higgs vev, which is rapidly changing during the phase transition. In order to estimate the parametric size of CP violation, we must first parameterize the $G$-axion mass $m_a$ as a function of temperature and Higgs vev. Assuming the dominant contribution to the axion mass comes from $G$-color confinement, this depends sensitively on the relative values of $m_Q, \Lambda_G,$ and the critical temperature $T_c$ of the electroweak phase transition.

In general, experimental constraints will restrict us to the case $m_Q > T_c$, leaving three possibilities:
\begin{enumerate}
\item $T_c < \Lambda_G < m_Q$, i.e., $G$-quarks above confinement, confinement before EWSB;
\item $T_c < m_Q < \Lambda_G$, i.e., $G$-quarks below confinement, confinement before EWSB;
\item $\Lambda_G < T_c < m_Q$, i.e., $G$-quarks above confinement, EWSB before confinement.
\end{enumerate}

The parametric dependence of the effective chemical potential, $\mu_{CS}$, or equivalently the change $\delta\Phi$ in the equation for the final baryon asymmetry, Eq.(\ref{eqn:baryonasymmetry}), is fairly sensitive to the hierarchy of scales.  Let us consider each in turn. \\

\begin{itemize}

\item {\bf $T_c < \Lambda_G < m_Q$}

Since in this case there are no light $G$-quarks at the confining scale, the axion mass may merely be estimated from instanton effects as
\beq
m_a^2 f_G^2 \sim \Lambda_G^4.
\eeq
Upon integrating out the massive $G$-quarks, however, the IR scale $\Lambda_G$ picks up a dependence on the masses $m_Q$ (for fixed value of the $G$-coupling-constant in the UV), and thus the Higgs vev $v(T)$, which at 1-loop is given by
\beq
\Lambda_G = \Lambda_{G,UV} \left(\frac{\Lambda_{G,UV}}{m_Q}\right)^{(b_{1,UV}/b_{1,IR} -1)}
\eeq
where $b_{1,UV}$ and $b_{1,IR}$ are the 1-loop beta-function coefficients of the $G$-color theory above and below the $G$-quark
mass threshold, and $\Lambda_{G,UV}$ is the $G$-color scale assuming $m_Q=0$.   In our case we desire the change in IR scale due to the
electroweak vev turning on, which arises due to the originally degenerate $G$-quark mass eigenstates of mass $\mu$ splitting, as in Eq.(\ref{eqn:mQT}). For the toy model, the confinement scale below the mass of all the $G$-quarks is given in terms of the $UV$ scale by
\beq
\Lambda_G = \Lambda_{G,UV} \left( \frac{\mu (\mu^2 - \frac{1}{2} \lambda \lambda' v^2)}{\Lambda_{G,UV}^3} \right)^{2/11N}.
\eeq 
In the case $T_c < \Lambda_G < m_Q$, we expect the ratio $v(T)/\mu$ to be small, so expanding the expression for $\Lambda_G^4$, we find that, for our toy model, the leading change at the electroweak phase transition is given by
\beq
\delta \Lambda_G^4 = -\frac{8 \lambda\lambda'}{11 N} \frac{v \delta v}{\mu^2} \Lambda_{G,UV}^{(4-24/11N)} \mu^{24/11N} .
\eeq
Similarly the leading change in the denominator $m_Q^4$ in Eq.(\ref{eqn:phidef}) is
\beq
\delta \sum_i \frac{1}{m_{Q,i}^4(T)} = 10 \lambda\lambda'  \frac{v \delta v}{\mu^6} .
\eeq
Putting these together we find that 
\beq
\delta \Phi =  \sin \theta_G   \left(10 - \frac{8}{11N}\right) \left(\frac{\lambda\lambda'  v \delta v}{\mu^2}\right) \left(\frac{\Lambda_{G,UV}}{\mu}\right)^{(4-24/11N)}, 
\label{eqn:deltaphicase1}
\eeq
Since $\delta v \leq v$, we see from Eq.(\ref{eqn:deltaphicase1}) that it is not possible to raise the $G$-quark mass scale arbitrarily above the EW-scale $v$ by raising the vector-like mass $\mu$, without strongly suppressing the final baryon asymmetry.

\item {\bf $T_c < m_Q < \Lambda_G$}
In this case the mass comes from confinement effects, with some number of light species of $G$-quark. The axion mass may thus be estimated using chiral perturbation theory, in analogy with the QCD axion, giving
\beq
m_a^2 f_G^2 \sim m_Q \Lambda_G^3.
\eeq
Dependence on the Higgs vev arises again through dependence on $m_Q$; in this case the result is simply
\beq
\delta \Phi \simeq  10 \sin \theta_G \lambda \lambda' v \delta v \frac{\Lambda_G^3}{\mu^5}.
\eeq
Ultimately, this will prove to be an uninteresting region of parameter space from the perspective of electroweak baryogenesis due to the rapid relaxation of $\theta_G$ when $\Lambda_G > m_Q$.

\item {\bf $\Lambda_G < T_c < m_Q$}

Of course, if $\Lambda_G < T_c$, the electroweak phase transition occurs before confinement. In this case, the axion still obtains a mass through instanton effects that may be estimated for large $T$ using the dilute-instanton-gas approximation \cite{Gross:1980br}. In this case, the axion mass scales as 
\beq
m_a^2 f_G^2 \approx \Lambda_G^4  \left(\frac{\Lambda_G}{T} \right)^{\frac{1}{3}(11 N  - 12)}. 
\eeq
The full calculation is fairly tedious; see the Appendix for details.
Once again, the $G$-quark mass enters into $\Lambda_G$ via scale-matching at the scale $m_Q$, and in the case of the toy model the axion mass may be expressed in terms of UV parameters as 
\beq
m_a^2 f_G^2 \approx \Lambda_{G,UV}^4 \left( \frac{\mu (\mu^2 - \frac{1}{2} \lambda \lambda' v^2)}{\Lambda_{G,UV}^3} \right)^{2/3} \left( \frac{\Lambda_{G,UV}}{T} \right)^{\frac{11 N}{3} - 4}.
\eeq
Then for our toy model we have parametrically in this case
\beq
\delta \Phi \simeq \frac{28}{3} \sin \theta_G \left( \frac{\lambda \lambda' v \delta v}{\mu^2} \right) \left( \frac{\Lambda_{G,UV}}{\mu}\right)^2 \left( \frac{\Lambda_{G,UV}}{T} \right)^{\frac{11 N}{3} - 4}.
\eeq

\end{itemize}

\begin{figure}[t!!]
\centering
\includegraphics[width=2.5in]{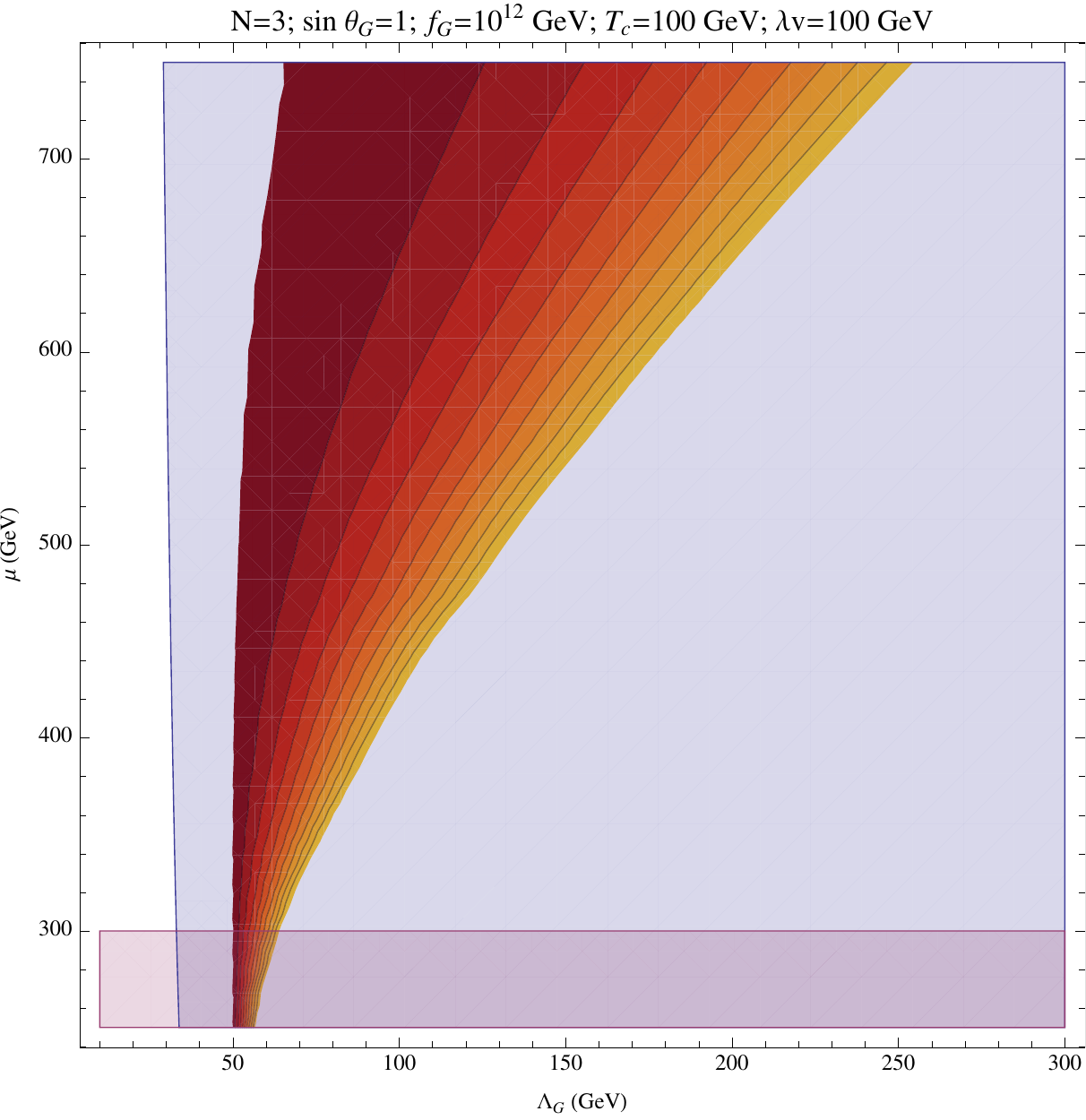} 
\includegraphics[width=2.5in]{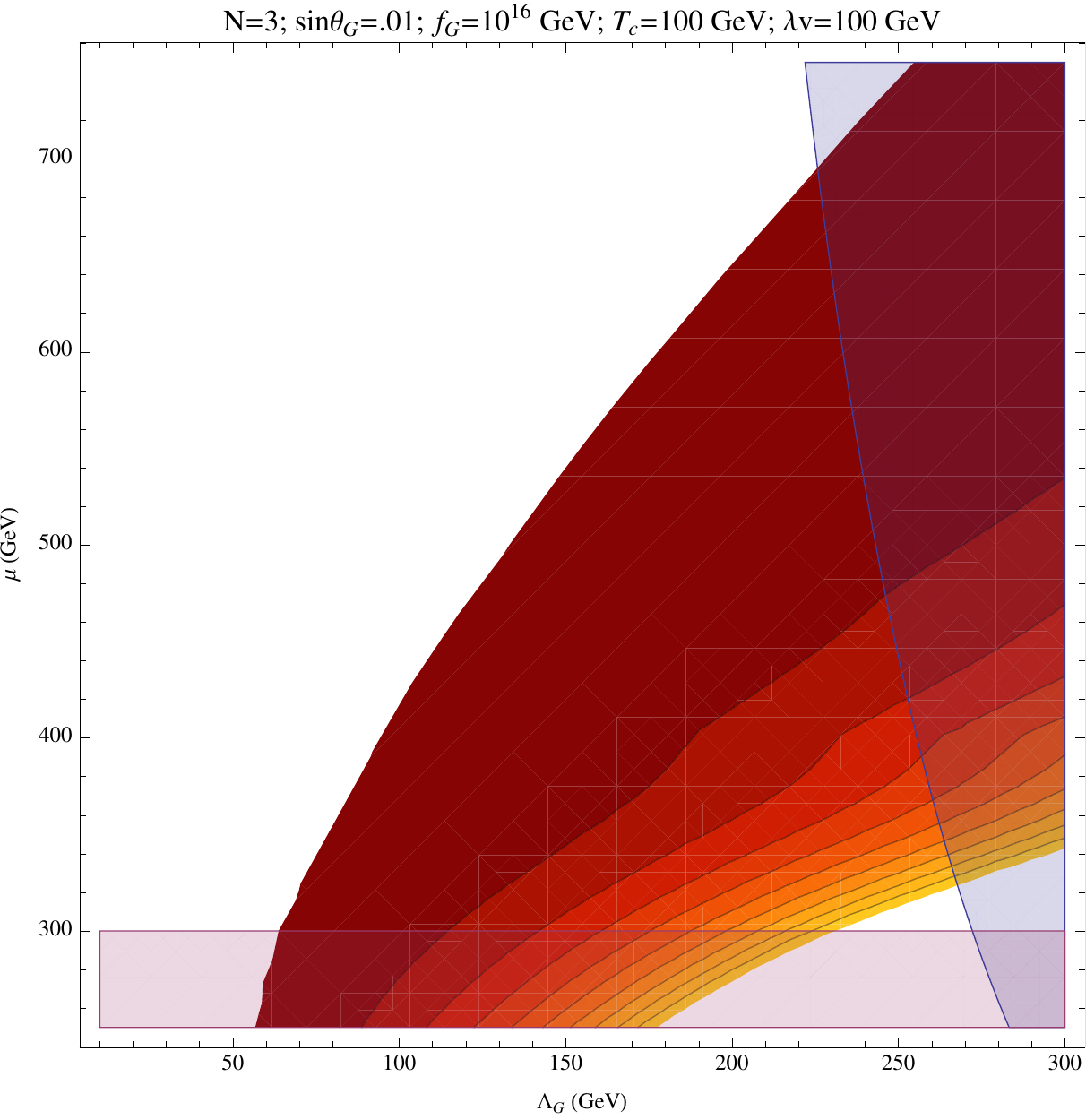} 
\caption{Baryon asymmetry as a function of $\Lambda_G, \mu$ for representative choices $N=3$, $T_c = 100$ GeV, $\sqrt{\frac{\lambda \lambda'}{2}} v = 100$ GeV,  and $\sin \theta_G = 1$, $f_G = 10^{12}$ GeV (left), $\sin \theta_G = 10^{-2},$ $f_G = 10^{16}$ GeV (right). We have taken a smooth interpolation of the various limiting values of $\delta \Phi$. The red contours indicate $10^{-11} < \Delta < 10^{-9}$ from dark to light. The region shaded in blue is excluded by relaxation of $\theta_G$ before the EWPT (see Appendix for details; this represents a fairly conservative constraint, and it may be possible for baryogenesis to be efficiently assisted even in this region). The region shaded in purple ($\mu < 300$ GeV) is excluded by collider limits on $\mu$ (see Section 4.1). We allow a large range of $\Delta$ to take account of the uncertainties introduced by our approximate treatment of the EWPT dynamics; the correlation between $\Lambda_G$ and $\mu$ could be made more precise given an explicit theory of the first-order EWPT dynamics and associated baryon number creation at the bubble-wall.   For hidden-axion decay constant, $f_a \lsim 10^{12}$ GeV, the initial axion misalignment angle $\theta_G$ can be large, while for the string-preferred value, $f_a \sim 10^{16}$ GeV, the misalignment must satisfy $\theta_G \lsim 10^{-2}$, as discussed in Section 4.3.2.}
\label{fig:contours}
\end{figure}

In all of these cases it is then a relatively straightforward matter to compute the baryon asymmetry $\Delta$ using Eq.(\ref{eqn:baryonasymmetry}). There is, of course, an important caveat, which is that the $G$-axion evolution has not relaxed $\theta_G$ to zero by the time of the electroweak phase transition. Approximately speaking, $\theta_G$ begins to relax when $m_a(T) \gtrsim 3 H(T)$; the equation of motion for $\theta_G$ is that of an underdamped oscillator with temperature-dependent angular frequency $m_a(T)$. In practice, the situation is somewhat more delicate, and requiring $m_a(T) = 3 H(T)$ at a temperature $T < T_c$ is overly conservative. Rather, we may study the time evolution of $\theta_G$ with a temperature-dependent axion mass and impose the reasonable constraint that $\theta_G$ not pass through zero before $T =T_c$; a detailed discussion is reserved for the Appendix.\footnote{It might be possible that significant evolution towards a small effective axion mis-alignment angle is acceptable, but there are concerns about the oscillating behavior of the axion substantially washing out the produced baryon asymmetry even in this case.} This results in a significant constraint on the $G$-color confinement scale even for an axion decay constant approaching the Planck scale. Indeed, for $f_G \sim10^{12}\gev$, requiring the axion to not be relaxing implies the scale $\Lambda_G \lsim \gev$, for which the final baryon asymmetry is insufficient for baryogenesis, thereby excluding our mechanism. The situation is much more favorable for GUT-scale axions with $f_G \sim 10^{16} \gev$, for which confinement scales as high as a few hundred GeV are allowed without relaxing $\theta_G$ before the EWPT. The implications of this are twofold: First, GUT-scale $G$-axions are strongly favored if this mechanism for baryogenesis is to be effective. Second, the $G$-color confinement scale must be near the weak scale -- and certainly not much higher -- in order to prevent the CP violating angle from relaxing too soon. This is extremely fortuitous from the perspective of collider phenomenology, as it forces the scales of the $G$-color sector to lie within reach of the LHC.

Representative values of $\Delta$ as a function of the vector mass $\mu$ and strong coupling scale $\Lambda_G$ are shown in Fig. \ref{fig:contours} along with constraints coming from the relaxation of $\theta_G$. As noted above, the constraints from relaxation of $\theta_G$ entirely exclude axions with low scales of PQ symmetry breaking, $f_G \lesssim 10^{12}$ GeV; in this case $\theta_G$ always begins oscillating before the electroweak phase transitions (though, again, this is a conservative criterion, and efficient baryogenesis may still be possible once $\theta_G$ has begun to evolve). The situation for GUT-scale PQ breaking is more favorable; $\theta_G$ relaxes after the EWPT as long as $\Lambda_G \lesssim 250$ GeV. For such values of $\Lambda_G$, the G-quark masses cannot be much more than $600$ GeV in order to produce sufficient baryon asymmetry. Thus the favored parameter space for both $\Lambda_G$ and $\mu$ is tightly constrained to lie around the weak scale. Although the precise value of the baryon asymmetry depends on the details of the electroweak phase transition and transport across bubble walls, there is significant room to reproduce a sufficient amount of CP violation if the $G$-quarks and confinement scale lie in the range GeV--TeV.

\section{Experimental Constraints \label{sec:constraints}}

The allowed parameter range of $G$-color confinement scale $\Lambda_G$ and $G$-quark masses $m_Q$ is constrained by a variety of considerations, including limits from nonobservation at the Tevatron; constraints from precision electroweak measurements; and cosmological limits on both $G$-color fields and the $G$-axion. Ultimately these considerations place lower bounds on $m_Q$ and $\Lambda_G$, as well as a relation between $f_G$ and $\theta_G$, but do not significantly impact the parameter space for axion-assisted baryogenesis.

\subsection{Collider constraints}

The constraints on fermions of the $G$-color sector coming from nonobservation at the Tevatron and LEP are not tremendously stringent, owing both to the lack of specific searches for hidden sector $G$-fermions and, in the models considered here, the absence of light $G$-quarks charged under $SU(3)_C$.

$G$-fermions may be pair produced at colliders through an off-shell photon, $Z$, $W$, or Higgs. The production rates for the first three processes are fixed entirely by the gauge couplings, while production through the Higgs depends on the size of the $\lambda, \lambda'$ Yukawa couplings as well as the Higgs mass. Their production at colliders is essentially that of the quirk scenario at large $\Lambda$ \cite{Kang:2008ea, Juknevich:2009ji, Juknevich:2009gg}: pair-produced $G$-fermions form a bound state connected by a $G$ flux tube, radiating kinetic energy and angular momentum through emission of photons, hadrons, and $G$ glueballs before recombining and annihilating into lighter SM states. For $G$-fermions produced via an off-shell photon or $Z$, the primary annihilation channel is into $G$ glueballs. $G$-fermions produced via an off-shell $W$  cannot annihilate into an SM charge-neutral state, and so instead annihilate into leptons or quarks via an off-shell $W$. 

Current collider limits are fairly weak. There are a variety of potential bounds coming from different channels at the Tevatron. Two primary channels giving rise to representative limits are
\begin{itemize}
\item $l + \gamma + \slash{E}_T$; CDF Run II limits on anomalous events involving a high-$p_T$ charged lepton and photon with missing transverse energy based on 929 pb$^{-1}$ of data \cite{Abulencia:2007zi} exclude masses for color-singlet $G$-fermions below 200 GeV; the bound for colored $G$-fermions is $\sim 250$ GeV.
\item $2 \gamma + \gamma$ and $2 \gamma + \tau$; CDF Run II limits on the inclusive production of diphoton events with a third photon based on 1155 pb$^{-1}$ \cite{Aaltonen:2009in} place a similar limit on $G$-fermion masses, excluding $\lesssim 200$ GeV. Limits from a diphoton plus tau search on 2014 pb$^{-1}$ place limits on $G$-quark masses below 250 GeV assuming $\lambda \sim 1$, which may be relaxed somewhat for smaller Yukawa couplings.
\end{itemize}
Taken together, these searches collectively limit $m_Q \gtrsim 200$ GeV for the lightest $G$-quark (or $m_Q \gtrsim 250$ GeV for colored $G$-quarks). Given that the lightest $G$-quark mass is $m_Q = \mu - \sqrt{\frac{\lambda \lambda'}{2}} v$, for $\sqrt{\frac{\lambda \lambda'}{2}} v \sim 100$ GeV this suggests a lower limit of $\mu \gtrsim 300$ GeV on the vector masses. It is important to emphasize that bounds on a SM fourth generation from Higgs searches at the Tevatron are vitiated in this case, since the $G$-fermions do not acquire their entire mass from electroweak symmetry breaking. 

\subsection{Precision electroweak constraints}

As with any theory involving additional states carrying electroweak quantum numbers and interactions violating the custodial symmetry of the Higgs sector, the $G$-color sector contributes to precision electroweak observables. If the only source of mass for $G$-color fermions arose from electroweak symmetry breaking, this would pose a stringent constraint on the matter content of the $G$-color sector. However, the addition of vector masses for $G$-color fermions allows them to be safely decoupled; vector masses above a few hundred GeV are more than sufficient to satisfy precision electroweak constraints. 

As usual, the contributions to precision electroweak observables may be quantified in terms of the $S$ and $T$ parameters \cite{Peskin:1991sw} (contributions to $U$ are negligible and weakly constrained). For $\Lambda_G > M_Z$, the appropriate degrees of freedom are composites of the strong $G$-color interactions, and a precise calculation of contributions to the $S$ and $T$ parameters is not possible. However, a reasonable estimate may be made on the basis of the one-loop contribution due to the fundamental fermions of the $G$-color sector.

\begin{figure}[t]
\centering
\includegraphics[width=2.5in]{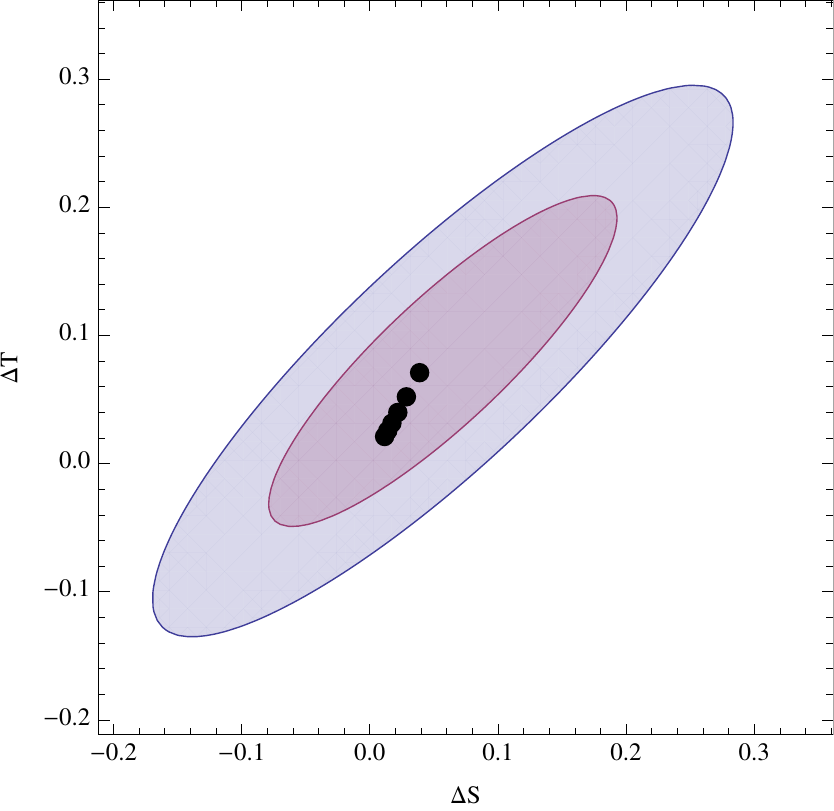} 
\includegraphics[width=2.5in]{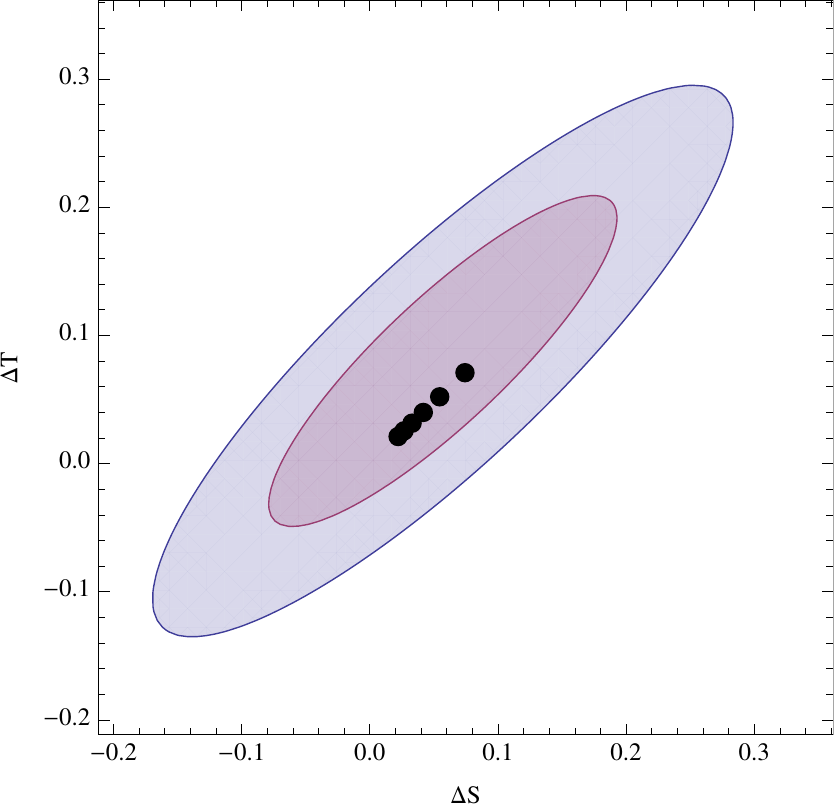} 
\caption{Corrections to electroweak precision observables $S,T$ for $\mu = 300,350,400,450,500,550$ GeV and $\sqrt{\lambda \lambda'/2} v = 100$ GeV assuming $G=SU(3)$ with hypercharge assignments $Y_Q = 0$ (left) and $Y_Q = 1/2$ (right). The $68\%,95\%$ CL ellipses are shown; best fit is achieved by $(\Delta S, \Delta T) = (0.057,0.080)$.}
\label{fig:pewc}
\end{figure}

For example, among the effective operators induced below the scale $m_Q$ there arises a contribution to the $S$ parameter of the form 
\beq
\lang_{eff} \sim  \frac{N g g'}{16 \pi^2} \frac{\lambda \lambda' \mu \mu'}{m_Q^4} H^\dag H W_{\mu \nu} B^{\mu \nu}
\eeq
It is clear from the level of the effective operator that this contribution to $S$ is suppressed by at least $(\lambda v / \mu)^2,$ which rapidly diminishes as a function of $\mu$. For $\mu = \mu'$ with an expansion in small $m_u \equiv \lambda v$, $m_d \equiv \lambda' v$, and $M_W$, the new fermion contributions to $S$ and $T$ may be computed more precisely using \cite{Martin:2009bg}:
\bea
\Delta T &=& \frac{N}{480 \pi s_W^2 M_W^2 \mu^2} \left[ 13 (m_u^4 + m_d^4) + 2 (m_u^3 m_d + m_d^3 m_u) + 18 m_u^2 m_d^2 \right] \\
\Delta S &=& \frac{N}{30 \pi \mu^2} \left[ 4 (m_u^2 + m_d^2) + m_u m_d (3 + 20 Y_{Q}) \right]
\eea
where $Y_Q$ is the weak hypercharge of $Q$ and $\Delta S, \Delta T$ indicate the deviations from the values of $S, T$ predicted by the SM alone for $m_t = 173.1$ GeV and $m_{h^0} = 115$ GeV. Of course, more exact estimates would need to take into consideration the effects of confinement, but the $(\lambda v / \mu)^2$ suppression will still persist as an effective form factor, and confinement is not expected to significantly alter the na\"{i}ve prediction and effects of decoupling. In general, precision electroweak constraints are readily satisfied for modest values of the vector mass $\mu$; representative points in the parameter space of $(\Delta S, \Delta T)$ are shown in Fig. \ref{fig:pewc}. The lower bound on $\mu$ coming from direct detection limits already renders safe the $G$-quark contributions to precision electroweak variables, even when the $SU(2)$ doublet $G$-quarks carry nonzero hypercharge.

\subsection{Cosmological constraints}

\subsubsection{$G$-color cosmology}

Naturally, one might be concerned that the addition of a strongly coupled gauge sector with couplings to the SM would pose a host of challenges to conventional cosmology. Although the safest cosmology might simply entail reheating temperatures $T_{RH} \lesssim$ MeV so that the $G$-color sector is never thermalized, such a low reheating temperature is fairly unattractive from the perspective of electroweak baryogenesis. Higher reheating temperatures lead to a cosmological abundance of $G$-quarks and $G$-color glueballs, whose evolution must be considered in detail. However, for the confinement scales of interest here, the $G$-color sector is surprisingly safe from a cosmological perspective.\footnote{Surprisingly, $G$-color sectors with much smaller $\Lambda_G$ -- as low as $\mathcal{O}(10 \text{ eV})$ -- are viable as well, though this often requires the lightest $G$-quarks to be charged under $SU(3)_C$ \cite{Jacoby:2007nw, Nussinov:2009hc}.}

As the $G$-quarks carry conserved $G$-color charges, the lightest $G$-quark is absolutely stable. However, below the confinement scale $\Lambda_G$, $G$-color interactions ensure that all $G$-color bound states annihilate efficiently into the lightest $G$-color mesons. The lightest $G$-color mesons then decay rapidly into SM states via their coupling to electroweak gauge bosons or the Higgs; for $\Lambda_G > 1$ GeV these decays happen well before BBN. 

For $\Lambda_G < m_Q$ the lightest states of the $G$-color sector tend to be $G$-color glueballs. These bound states decay into SM states via loops of $G$-color quarks, which then decay into SM states via the Higgs or electroweak gauge bosons. The leading dimension-six operator comes from Higgs boson exchange via the operator
\beq
\lang \supset \frac{\alpha_G \lambda^2}{3 \pi m_Q^2} H^\dag H G_{\mu \nu} G^{\mu \nu}
\eeq
which leads to a lifetime for the scalar $0^{++}$ $G$-color glueball of order \cite{Juknevich:2009gg}
\beq
\tau \sim 10^{-18} \text{ s} \times \left( \frac{m_Q}{300 \text{ GeV}} \right)^4 \left( \frac{100\text{ GeV}}{\Lambda_G} \right)^{7}.
\eeq
Needless to say, such decays occur well before BBN and pose no particular challenges to conventional cosmology.

\subsubsection{$G$-axion cosmology\label{sec:axioncosmo}}

Although the $G$-quarks and $G$-color glueballs are cosmologically safe due to the relatively large values of $m_Q$ and $\Lambda_G$, the cosmology of the light $G$-axion -- as with the cosmology of the QCD axion \cite{Preskill:1982cy, Abbott:1982af, Dine:1982ah} -- is rather less trivial. As with the QCD axion, the $G$-axion is initially a random field on superhorizon scales with vacuum expectation value $a_G \sim f_G \theta_{G,i}$; the fluctuations of $\theta_{G,i}$ are naturally expected to be of order one. Not far above the $G$-color confinement scale $\Lambda_G$, $G$-color instantons (and perhaps also string instantons) induce a potential for the $G$-axion. When $m_a \sim 3 H$, the axion begins to oscillate around the resulting minimum as an underdamped harmonic oscillator of frequency $m_a$. These coherent axion oscillations contribute as cold dark matter to the critical energy density an amount
\beq
\Omega_a h^2 \sim 10^7 \left( \frac{f_G}{M_P} \right) \left( \frac{\Lambda_G}{T_i}\right) \left( \frac{a_{G,i}}{f_G} \right)^2
\eeq
In this case $T_i$ is set by $m_a(T_i) \sim 3H.$ Consequently, the observed energy density constrains $f_G \left( \frac{a_{G,i}}{f_G} \right)^2 \lesssim 10^{12}$ GeV. If $\frac{a_{G,i}}{f_G} = \theta_{G,i} \sim 1$, then necessarily $f_G \lesssim 10^{12}$ GeV. However, this should not be construed as ruling out higher values of $f_G$; indeed, it is entirely possible for larger values of $f_G$ that the initial angle $\theta_{G,i}$ is anthropically constrained to satisfy the critical density bound \cite{Tegmark:2005dy}. For example, for a $G$-color PQ scale of order $f_G \sim 10^{16}\gev$, the critical density bound implies $\theta_{G,i} \lesssim 10^{-2}$.

\section{Phenomenology \label{sec:pheno}}

Ultimately, generating the observed baryon asymmetry from the strong CP angle of a confining hidden sector is most interesting if the hidden sector degrees of freedom are experimentally accessible. As we have seen in previous sections, the observed value of baryon asymmetry is most readily produced if the $G$-quark masses lie in the range 200 -- 600 GeV and the confinement scale lies in the range 50 -- 250 GeV , an ideal scenario for production at the LHC.

\subsection{LHC signatures}

If the $G$-axion accounts for the bulk of CP violation during baryogenesis, it is necessarily the case that the $G$-color confinement scale $\Lambda_G$ and $G$-quark masses $m_Q$ are in a range amenable to extensive production at the LHC. Indeed, effective axion-assisted baryogenesis provides a ``reason'' for the hidden sector to lie near the weak scale. The LHC phenomenology of an additional strong sector is quite striking, and has been studied extensively in \cite{Burdman:2008ek, Juknevich:2009ji, Juknevich:2009gg}; here we will confine ourselves to a brief review of the most compelling results.

For $m_Q, \Lambda_G$ in the range 100 GeV $-$ 1 TeV, the production and decays of $G$-color fields are prompt, occur within the detector, and may lead to distinctive signatures. As discussed earlier, CDF limits constrain the lightest $G$-quark to satisfy $m_Q \gtrsim 200$ GeV. $G$-quark pairs in the range $200 \text{ GeV} < m_Q <$ TeV may be directly produced at the LHC via off-shell electroweak gauge bosons or the Higgs. For $m_Q > \Lambda_G$, the primary decays of $G$-quark bound states are into $G$-color glueballs, whose subsequent decays into SM states are visible for $\Lambda_G > 1$ GeV. $G$-quark bound states carrying electroweak quantum numbers decay into $G$-color glueballs and SM states via an off-shell $W$, which may provide a particularly promising signal at the LHC \cite{Burdman:2008ek}. 

The decays of $G$-color glueballs are fairly spectacular, occuring primarily into electroweak bosons as well as the Higgs. The final states of these decays are rich in jets, leptons, and photons, with the most likely discovery channels coming from two- or four-photon resonances. In the parameter range of interest, $\Lambda_G \gtrsim 50$ GeV, the distinctively stringy signatures of the original quirk scenario \cite{Kang:2008ea} are less apparent due to the high scale of confinement, but instead the decays occur promptly in the detector and are by no means uninteresting.

Of course, definitively connecting the detection of a confining hidden sector at the LHC to axion-assisted electroweak baryogenesis is a much more delicate issue. Although the observation of $G$-quarks and $G$-color glueballs would not be an unequivocal indicator, their presence near the weak scale would be a compelling suggestion that hidden sector physics may be relevant to electroweak baryogenesis. Further evidence would be provided if the parameters of new physics indicated that the electroweak phase transition may have been sufficiently strongly first-order. For example, in the case of the Minimal Supersymmetric SM (MSSM), a Higgs mass below 127 GeV and stop mass below 120 GeV would be suggestive of a first-order electroweak phase transition sufficient for electroweak baryogenesis \cite{Carena:2008vj}.

\subsection{Axion phenomenology} 

As $f_G \lsim M_{Pl}$, the $G$-axion can be no lighter than $m_a \sim 10^{-7}$ eV. This puts it outside the range of the interesting gravitational signatures of ultra-light axions considered in \cite{Arvanitaki:2009fg}. The $G$-axion does develop a coupling to $\vec E \cdot \vec B$ at two loops, raising the prospect of interesting signatures from decays into photon pairs, but in general this rate is far too small to be of observable interest. For the most part, the prospects for detecting the $G$-axion through direct or indirect means are not terribly better or worse than those for the QCD axion.

It should be noted that the effective $\theta$ angle for $SU(2)$ induced by Eq.(\ref{eqn:su2theta}) is invisible to measurements in the current era irrespective of the action of the $G$-axion, owing to the fact that a time-independent weak $\theta$ angle may be eliminated by $(B+L)$ transformations; even in the presence of $(B+L)$-violating irrelevant operators, the effective $\theta$ angle is highly suppressed. Only the time-dependent angle during baryogenesis is of any relevance to phenomenology in this case.

\subsection{Strengthening the phase transition}

It is amusing to note that the addition of weak-scale fermions coupled to the Higgs may significantly increase the strength of the electroweak phase transition \cite{Carena:2004ha}. In general, the effects are greatest for $\lambda v \gg \mu$. Making the SM phase transition truly first order requires (for a moderate number of $G$-color degrees of freedom) strong coupling to the Higgs of order $\lambda \gtrsim 1.5$; such strong couplings raise the unpleasant prospect of Yukawa coupling Landau poles several decades above the weak scale. It is clear that, given the range of $\lambda$ and $\mu$ considered here, the $G$-color sector cannot render the phase transition strongly first-order on its own. Nonetheless, the presence of $G$-fermions coupled to the Higgs certainly increases the strength of the phase transition. Such effects may be even more significant in the supersymmetric case, where the MSSM phase transition may already be weakly first order in some regions of parameter space. Indeed, they may even raise the bounds on Higgs and stop masses required for a first-order electroweak phase transition above $\sim 120$ GeV.

\section{Supersymmetric Axion-assisted Electroweak Baryogenesis \label{sec:susy}}

Although we have been working thus far in a nonsupersymmetric context, it is fairly straightforward to generalize our results from the SM to the MSSM. In this case the fermions $Q, \overline Q, U, \overline U$ may be promoted to superfields with superpotential terms
\beq
W_G = \mu_Q Q \overline Q + \mu_U U \overline U + \lambda H_u  Q \overline U + \lambda' H_d \overline Q U
\eeq
Assuming the squarks and sleptons are heavier than a few hundred GeV, the existing constraints from colliders, cosmology, and precision electroweak considered in previous sections are essentially unaltered. 

It is important to emphasize that the vector masses $\mu_Q, \mu_U$ may be generated by the same physics that gives rise to the supersymmetric Higgs sector $\mu$-term $\mu_H H_u H_d$, and hence naturally lie around the weak scale. This provides a natural explanation for the coincidence problem between the scale of $G$-color and SM sectors.

Assuming unbroken $R$ parity, the lightest supersymmetric particle (LSP) will be stable and may contribute to dark matter relic abundance. The LSP in a supersymmetric theory depends on the scale of supersymmetry breaking and its communication to both the $G$-color sector and the MSSM. One potential candidate for a $G$-color-sector LSP is the $G$-color gluino, leading to the formation of $G$-color gluino-gluon hadrons. Another candidate is the $G$-color axino, the fermionic superpartner of the $G$-axion. In both cases, the relic abundance is in general incalculable owing to the strong $G$-color interactions, but both hidden sector LSP candidates may lead to consistent cosmologies and dark matter abundances. In the event that the LSP arises in the $G$-color sector, MSSM LOSP decays into the hidden sector LSP are expected to give rise to the usual host of interesting hidden valley signals. 

\subsection{Little hierarchy problem}

It has been observed \cite{Martin:2009bg, Graham:2009gy} that a vector-like fourth generation carrying SM quantum numbers and coupling to the Higgs may ameliorate the little hierarchy problem of the MSSM by virtue of their additional radiative contributions to the lightest Higgs mass. Strictly speaking, nothing requires this generation to carry SM quantum numbers alone; $G$-quarks carrying electroweak quantum numbers and coupling to the Higgs may play the same role as a vector-like fourth generation. Crucially, $G$-color quark loop contributions to the lightest Higgs mass will be $N$ enhanced, akin to the color enhancement of fourth-generation vector quarks.  For moderate values of $\tan \beta$ and $\lambda \sim \mathcal{O}(1)$, vector masses in the range 300 GeV $< \mu <$ 550 GeV may suffice to naturally push the Higgs mass above 114 GeV. Conveniently, this mass range is ideally suited to generating the observed baryon asymmetry and distinctive $G$-color phenomenology at the LHC. 
We find this amusing.   

\section{Conclusions \label{sec:conc}}

In this work we have explored a mechanism to generate the additional CP violation required for successful EW baryogenesis.  If there exists a hidden gauge sector with a strong coupling scale close to the electroweak scale, then the CP-violating strong dynamics of the hidden sector $\theta$-term may feed into the SM sector and induce an effective chemical potential for baryon number in the early universe, allowing non-perturbative (``sphaleron'') processes in the SM to generate the observed baryon asymmetry.  Communication of CP violation to the SM requires messenger states charged under both the confining hidden sector gauge group and, at least, $SU(2)\times U(1)_Y$.  The mechanism is efficient when the hidden strong scale is at or slightly above the electroweak scale, and the messenger states are also parametrically close to the electroweak scale. These scales are bounded from below by collider searches and bounded from above by the relaxation of the hidden sector $\theta$ term.  In this case the effective CP-violating chemical potential in the SM is large precisely due to the combination of the strong dynamics of the hidden sector and the fast (compared to the Hubble time at electroweak temperatures) dynamics of the EW phase transition.  Thus, for a sufficiently out-of-equilibrium EW phase transition, large CP violation can easily be induced by our mechanism, independent of the CP violation in the SM, or MSSM, sector.  At temperatures parametrically below the electroweak scale the CP violation turns off due to the relaxation of the hidden $\theta$-angle by the hidden sector axion (which is heavier than the QCD axion assuming equal decay constants). 

Simple supersymmetric and non-supersymmetric toy realisations of this mechanism involving $SU(2)_L \times U(1)_Y$-charged vector-like states also charged under the hidden group were presented.   The hidden sector, including the matter messengers in particular, easily evade current experimental and observational constraints, but the messengers are experimentally accessible at the LHC over the preferred parameter space.  The collider phenomenology of the messengers and hidden sector gauge fields has much in common with ``hidden valley'' or ``quirk'' scenarios, while the connection of the hidden sector and messenger mass scales with the EW scale via successful axion-assisted EW baryogenesis provides a reason why such new, hidden-valley, physics sits parametrically
close to the weak scale.  Intriguingly, in the supersymmetric case the messenger matter naturally solves the little hierarchy problem of the MSSM when the messengers are at a mass scale that allows efficient axion-assisted EW baryogenesis.   Finally, our axion assistence
mechanism favours the natural string value of the axion decay constant, $f_G \sim 10^{16} \gev$, implying that the hidden axion
should comprise a significant fraction of the observed dark matter density.

\subsection*{Acknowledgements}
We would like to thank Savas Dimopoulos, Sergei Dubovsky, Peter Graham, Juan Garcia-Bellido, Lawrence Hall, and Jay Wacker for useful discussions. JMR thanks the Stanford Institute for Theoretical Physics and the Berkeley Center for Theoretical Physics for their kind hospitality during a Visiting Professorship
at Stanford during the initiation of this work; NC thanks the Dalitz Institute for Fundamental Physics and the Department of Theoretical Physics, Oxford University for hospitality during the completion of this work. NC is supported by the NSF GRFP and the Stanford Institute for Theoretical Physics under NSF Grant 0756174.   JMR is partially supported by the EU FP6 Marie Curie Research and Training Network UniverseNet (MPRN-CT-2006-035863),  by the STFC (UK), and by a Royal Society Wolfson Merit Award.  

\appendix
\section{Axion mass and evolution}
At finite temperature, the axion mass may be determined (in the limit $T \gg \Lambda_G$) by integrating over instantons of all sizes, $m_a^2 f_G^2 = \int d \rho \, n(\rho)$, where $\rho$ is the instanton size and $n(\rho)$ the number density. Using the dilute-instanton-gas approximation \cite{Gross:1980br}, one finds the zero-temperature result 
\beq
n(\rho, T =0) = \frac{C_N}{\rho^5} \left( \frac{4 \pi}{g^2} \right)^{2N} \left( \prod_i^{N_f} \xi \rho m_i \right) \exp \left( - 8 \pi^2 g^2 \right)
\eeq
where where $\xi = 1.33876$, $C_N = \frac{(0.260156) \xi^{-(N-2)}}{(N-1)! (N-2)!}$, and $g$ is given by the renormalization group improved running coupling at the scale $\rho$,
\beq
\frac{4 \pi^2}{g^2(\rho)} = \frac{1}{6} (11 N - 2 N_f) \ln (1/\rho \Lambda) + \frac{1}{2} \frac{[17 N^2 - N_f (13 N^2 - 3)/2 N]}{(11 N - 2 N_f)} \ln \ln(1/\rho \Lambda) + \mathcal{O}(1/\ln \rho \Lambda)
\eeq 
The finite-temperature density is related by 
\beq
n(\rho, T) = n(\rho, 0) \times  \exp \left(- \left \{ \frac{1}{3} \lambda^2 (2 N + N_f) + 12 A(\lambda) [1 + \frac{1}{6} (N - N_f)] \right \} \right) 
\eeq
where  $\lambda = \pi \rho T$, $A(\lambda) \simeq - \frac{1}{12} \ln (1 + \lambda^2/3) + \alpha (1 + \gamma \lambda^{-3/2})^{-8}$, $\alpha = 0.012897$, and $\gamma = 0.15858$.
Using this, we may derive an expression for the finite-temperature axion mass, valid for $T \gg \Lambda_G$:
\beq
m_a(T)^2 = \frac{\Lambda_G^4}{f_G^2} \left( \frac{m_Q}{\Lambda_G} \right)^{N_f} \left( \frac{\Lambda_G}{T} \right)^{\frac{1}{3} (11 N + N_f - 12)} \mathcal{I}
\eeq
where
\bea
\mathcal{I} &\equiv& \frac{0.260156}{(N-1)!(N-2)!} \left( \frac{11 N -2 N_f}{6} \right)^{2N} \xi^{N_f - N+2} \\ \nonumber
&\times& \int_0^\infty x^{\frac{1}{3}(11 N + N_f - 15)} \ln\left(T/x\Lambda_G \right)^{2N-a} \exp[f(x)] \, dx 
\eea
Here
\beq
f(x) \equiv - \frac{1}{3} \pi^2 x^2 (2 N + N_f) + \frac{1}{6} \left(6 + N - N_f \right) \left[\ln(1+\pi^2 x^2/3)-12 \alpha (1+ \gamma/\pi^{3/2} x^{3/2})^{-8}\right]
\eeq
and $a = \frac{[17 N^2 - N_f (13 N^2 - 3)/2N]}{(11 N - 2 N_f)}$. This expression gives the correct finite-temperature axion mass in the limit $m_Q \gg \Lambda_G$ by taking $N_f \to 0$ (although in this case $\Lambda_G$ must then be related to $\Lambda_{G,UV}$ via the usual scale-matching). The instanton density $n(\rho, T=0)$ may also be used to obtain an estimate of the zero-temperature axion mass, but here the issue is more delicate. While at finite temperature it is possible to integrate over instantons of all sizes because $T$ provides the appropriate large-distance cutoff, at zero temperature there is no such cutoff. At best, one may estimate the zero-temperature mass by assuming that instantons of size $\rho \sim \frac{1}{\Lambda_G}$ dominate, which gives the parametrically correct result.

\begin{figure}[t]
\centering
\includegraphics[width=4in]{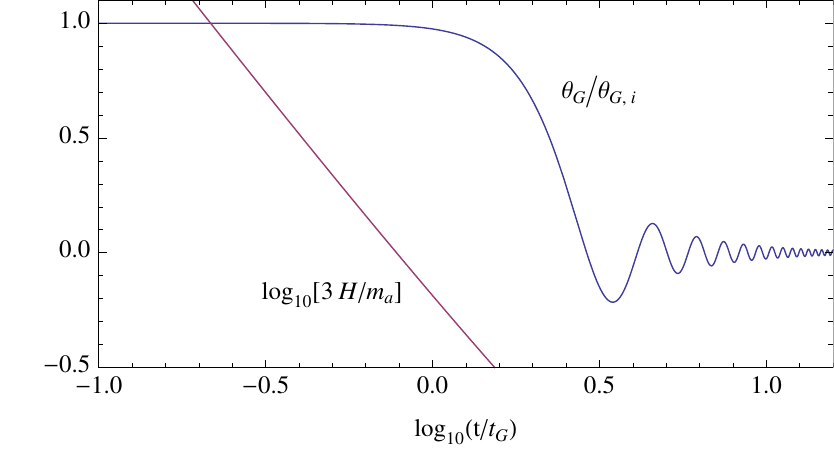} 

\caption{The evolution of $\theta_G$ as a function of time (where $t_G = t(T=\Lambda_G)$) for $\mu = 300$ GeV, $\Lambda_G = 400$ GeV, $f_G = 10^{16}$ GeV, where we have taken for $m_a(T)$ a smooth interpolation between $m_a(T \gg \Lambda_G)$ and $m_a(T=0)$.   The straight line denotes $\log_{10}[3 H(T)/m_a(T)]$. Note that $\theta_G$ begins to relax well after $3 H(T) = m_a(T)$. }
\label{fig:evo}
\end{figure}

Having determined the axion mass for (at least for $T = 0$ and $T \gg \Lambda_G$), we may now turn to the evolution of $\theta_G$. This is of critical importance in determining the scale at which CP violation from the $G$-color sector begins to decrease. The equation of motion for $ \theta_G$ is simply
\beq
\ddot{\theta}_G + 3 H \dot \theta_G + m_a^2(T) \theta_G = 0
\eeq
(where we have disregarded the negligible $\Gamma_a \dot \theta_G$ term). This evolution becomes interesting thanks to the temperature dependence of the axion mass, which becomes important when $m_a(T) \gtrsim 3H$, at which point coherent oscillations of $\theta_G$ commence. We may solve the equations of motion numerically, assuming a functional form for $m_a(T)$ that smoothly interpolates between $m_a(T \gg \Lambda_G)$ and $m_a(T=0).$ A representative evolutionary history for $\theta_G$ is shown in Figure~\ref{fig:evo}. There are two details worth emphasizing: The first is that $\theta_G$ begins to evolve significantly after the point $3 H(T) = m_a(T)$ is reached, by as much as an order of magnitude. The second is that this evolution commences after $t_G \equiv t(T= \Lambda_G)$, i.e., {\it after} the confining phase transition, despite the fact that $3 H(T) = m_a(T)$ occurs {\it before} $t_G$. Both considerations become significant when determining constraints on axion-assisted electroweak baryogenesis coming from the relaxation of $\theta_G$.

\bibliographystyle{JHEP}
\bibliography{axibaryrefs}

\end{document}